\documentclass[conference]{IEEEtran}

\usepackage{cite}
\ifCLASSINFOpdf
   \usepackage[pdftex]{graphicx}
   \DeclareGraphicsExtensions{.pdf,.jpeg,.png}
\else
 \fi

\usepackage[cmex10]{amsmath}

\usepackage{subfig}
\usepackage{tabularx}

\usepackage{fixltx2e}
\DeclareMathSizes{8}{8}{8}{8}

\begin{document}

\title{Development and accuracy evaluation of Coded Phase-shift 3D scanner}

\author{\IEEEauthorblockN{Pranav Kant Gaur}
\IEEEauthorblockA{Computer Division\\Bhabha Atomic Research Centre\\Mumbai,INDIA\\
Email: pranav@barc.gov.in}
\and
\IEEEauthorblockN{D.M.Sarode}
\IEEEauthorblockA{Computer Division\\Bhabha Atomic Research Centre\\Mumbai,INDIA\\
Email: dinesh@barc.gov.in}
\and
\IEEEauthorblockN{S.K.Bose}
\IEEEauthorblockA{Computer Division\\Bhabha Atomic Research Centre\\Mumbai,INDIA\\
Email: bose@barc.gov.in}}

\maketitle

\begin{abstract}
In this paper, we provide an overview of development of a structured light 3D-scanner based on combination of binary-coded patterns and sinusoidal phase-shifted fringe patterns called Coded Phase-shift technique. Further, we describe the experiments performed to evaluate measurement accuracy and precision of the developed system. A study of this kind is expected to be helpful in understanding the basic working of current structured-light 3D scanners and the approaches followed for their performance assessment.
\end{abstract}

\IEEEpeerreviewmaketitle

\section{Introduction}
\IEEEPARstart
Phase-shift approach[1][2] encodes the projector pixel coordinates with
the phase of projected sinusoidal fringe patterns. When this
pattern is projected onto the scene of interest it codifies the
scene as well using the phase of the incident sinusoidal
signal. Camera captures the scene with projected patterns.
Recovery of phase information at all captured points in the
scene provides information regarding source projector
pixel(using phase value). Hence stereo-correspondence
between camera and projector pixels can be established. But
since the sinusoidal patterns are periodic in nature, the phase
value repeats after every $2\pi$(or a period) therefore binary coded patterns[3]
are used in addition to assist in recovering original phase at
each point in the captured scene by assigning unique period
number to each cycle of sinusoidal fringe[4]. This approach
eliminates the ambiguity due to periodicity of sinusoidal
fringes. \newline
\indent This combined approach has the advantage of high
accuracy of phase-shift technique and high robustness
against noise as of binary-coded pattern based technique[3].\newline
\indent Development work to realize this approach was initiated
because in our knowledge there is no Open-source
implementation of this technique which can be used as a
starting point for accurate 3D measurements. Authors in [5-6] have
developed such systems but in [5] there are no provisions for
system calibration and the work found in [6] do not have documentation and is
developed only for Windows platform.\newline
\indent Furthermore, the performance of a metrology equipment has been conventionally measured in terms of its measurement accuracy and precision(or repeatability). Recently, there have been many attempts to evaluate performance of another
popular 3D sensor Microsoft Kinect[7-10]. Same methods
can be applied to evaluate performance of our developed
system. But these approaches assess measurement accuracy
either with respect to 3D data obtained from laser scanner[7]
which requires accurate calibration of laser scanner itself, or
confirm to VDI/VDE 2634 standard which requires accurate
fabrication of sphere balls and hexagonal structures which
again requires sophisticated fabrication[9]. To avoid such
overheads, this paper reports a simple and straightforward method for assessing
measurement accuracy. Further, we report the precision of
the developed system.\newline 
\indent In section II, process for estimating stereo-
correspondence using Coded phase-shift technique is
described followed by description of process of system
calibration and finally, triangulation. In section III, approach used for evaluation of measurement accuracy and precision has been described. Section IV concludes the paper.

\section{Development}
A structured-light 3D scanner based on optical
triangulation(shown in Figure 1) needs to determine the
correspondence between camera and projector pixels. This
goal is achieved by projection of a known pattern on object
of interest. This pattern(e.g. light stripes in Figure 1) assign a
unique code to each point on the surface of object. Camera
captures a view of scene and recovers this code at each pixel.
This process indirectly relates a projector column(\textit{strip
number} in Figure 1) or row or a unique combination of row and
column to a camera pixel.\newline
\indent Once the correspondence is known, equations for optical
rays emanating from any corresponding pair of camera and
projector pixels needs to be known in real world units, for
which system calibration is performed.\newline
\indent Given stereo-correspondence and system calibration
information, optical triangulation can be performed to
compute 3D coordinates for each real-world point seen by both
camera and projector.\newline

\begin{figure}[ht]
\centering
\includegraphics[width=7cm,height=5cm]{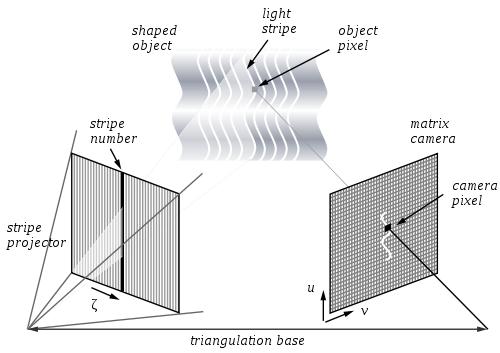}
\hspace{5cm}\caption{Optical triangulation(Source:Wikipedia)}
\end{figure}

\subsection{System setup}
In this work, we have used Logitech Quickcam Sphere AF
webcam at 1600X1200 resolution, Sharp PG-F200X
projector at 1024X768 resolution. Figure 2 shows our system setup.
\begin{figure}
\centering
\includegraphics[width=7cm,height=5cm]{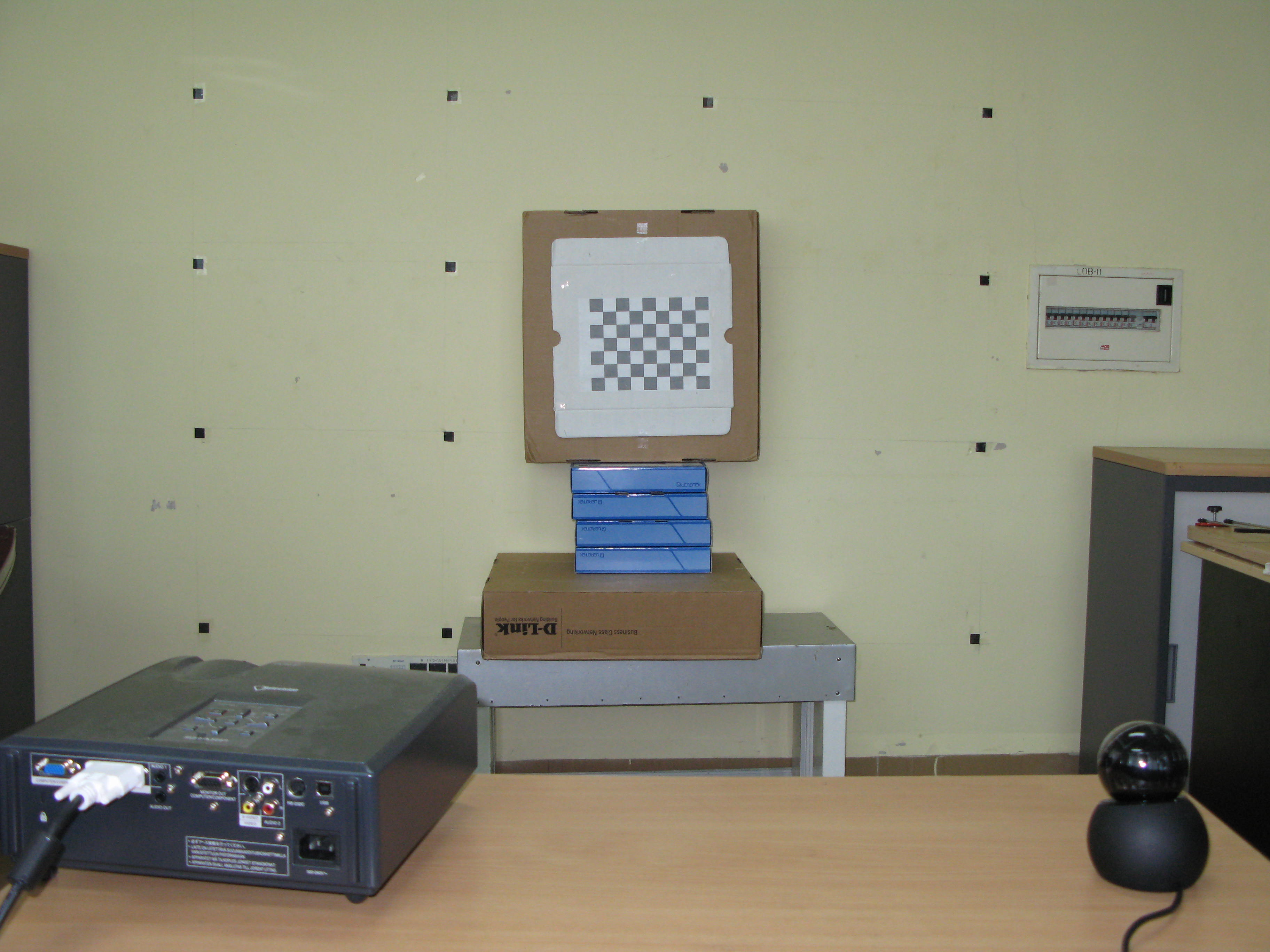}
\caption{System setup}
\end{figure}

\subsection{System calibration}
Camera and projector calibration is a process of estimating their intrinsic(i.e. focal length:($f_x,f_y$), principle-point:($c_x,c_y$)) and extrinsic geometry(i.e. rotation:($r_x,r_y,r_z$), translation:($t_x,t_y,t_z$))[11-13]. \newline
\indent In this work we have used OpenCV library for camera calibration and pose estimation
algorithms for intrinsic and extrinsic calibration respectively. For projector intrinsic calibration we have used VPCLib[14] since we have experimentally observed unacceptably lower repeatability of OpenCV calibration algorithm when used for projector.

\subsubsection{Camera calibration}
A checkerboard of known dimension is shown at different
distances and orientations with respect to camera. Figure 10
shows the plot of the views used for camera calibration. This
process gives sample 2D image point and corresponding 3D
points. This information is used to estimate calibration
parameters using OpenCV method. Table I shows the
estimated calibration parameters for camera.

\subsubsection{Projector calibration}
Projector can be modeled as inverse-camera[15]. Projector
calibration is performed with camera as a feedback device.
Initially, camera-to-screen(or any planerboard)homography(projective-mapping between 2 planes) is
computed. Then projector projects the checkerboard pattern.
Camera captures the projected checkerboard pattern and
detects its inner-corners. Using camera-to-screen
homography world coordinates for detected checkerboard
corners are computed. This process gives set of 2D-3D
correspondences required for projector calibration.
Thereafter, an identical procedure as of camera calibration
can be applied for estimating projector calibration
parameters. Figure 11 shows plot of the views used for
projector calibration. Table I shows estimated calibration
parameters for projector.

\subsubsection{Extrinsic camera-projector calibration}
To bring optical-ray from a camera-pixel and optical-ray
emanating from corresponding projector-pixel to a common
coordinate system, relative rotation and translation between
camera and projector coordinate system needs to be known.
This is required for performing optical triangulation since
computed 3D-coordinate is the intersection point of optical
rays from camera and projector, requiring both rays to be in
a common coordinate system.\newline
\indent Since intrinsic parameters are already known for both
projector and camera, similar procedure as used for
camera(and projector) calibration is applied but for a single
view of physical(and projected) checkerboard to get 2D-3D
mappings for camera and projector separately. These
mapping are used to estimate rotation and translation
transformation for camera and projector coordinate systems
with respect to world coordinate system. These are then
combined to get projector-to-camera rotation and translation
parameters.

\begin{table}[!t]
\caption{Camera and Projector intrinsic calibration parameters}
\centering
\begin{tabular}{c c c}
\hline
Parameter & Camera & Projector\\
\hline
$f_x$ & 1362.2 & 2261.7 \\
$f_y$ & 1372.2 & 2262.8 \\
$c_x$ & 803.9 & 522.7 \\
$c_y$ & 590.1 & 713.8 \\
$k_1$ & 0.07  &  0.0 \\
$k_2$ & -0.14 & 0.0 \\
\hline
\end{tabular}
\end{table}

\subsection{Stereo correspondence}
Once system intrinsic and extrinsic geometry is defined,
stereo-correspondence which pairs the points in camera and
projector viewing a common 3D point is estimated.
Following subsections describe the working of modules for
estimating stereo-correspondence.

\subsubsection{Pattern generation module}
This module generates phase-shifted sinusoidal fringes and binary-
coded patterns. Equation 1 shows the relation used for generating the phase-shifted sinusoidal fringes.

\begin{equation}
\begin{aligned}
& I_1=I_{dc}+I_{mod}*cos(\phi+\theta) \\
& I_2=I_{dc}+I_{mod}*cos(\phi) \\
& I_3=I_{dc}+I_{mod}*cos(\phi-\theta) \\
\end{aligned}
\end{equation}
where, $I_1$,$I_2$,$I_3$ represent 3 sinusoidal signals at any point
represented by phase $\phi$ and successively shifted in phase by
$\theta$. $I_{dc}$ models a biasing factor, $I_{mod}$ models the modulation
intensity of the sinusoid. Further, binary-coded patterns are
designed such that width of a bit-plane equals the width of a
fringe(or one sinusoidal cycle).

\subsubsection{Pattern projection and capture module}
This module sequentially projects and captures the sinusoidal
phase-shifted and binary-coded patterns. Figure 3 shows
some captured vertical and horizontal fringe and binary
coded pattern images.

\begin{figure}
\centering  
\def\tabularxcolumn#1{m{#1}}  
\begin{tabularx}{\linewidth}{@{}cXX@{}}  
\begin{tabular}{c c}  
\subfloat[]{\includegraphics[width=4cm,height=4cm]{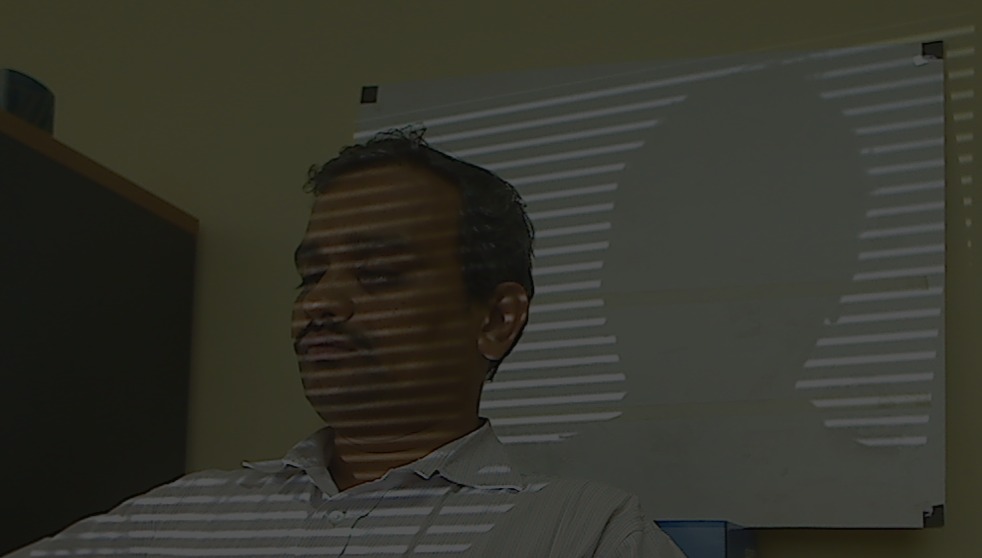}} &
\subfloat[]{\includegraphics[width=4cm,height=4cm]{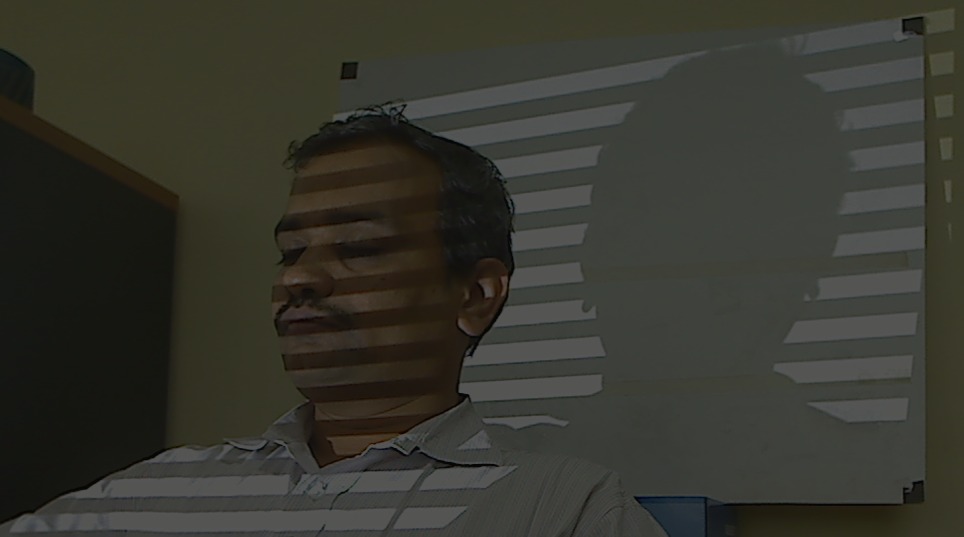}} \\
\subfloat[]{\includegraphics[width=4cm,height=4cm]{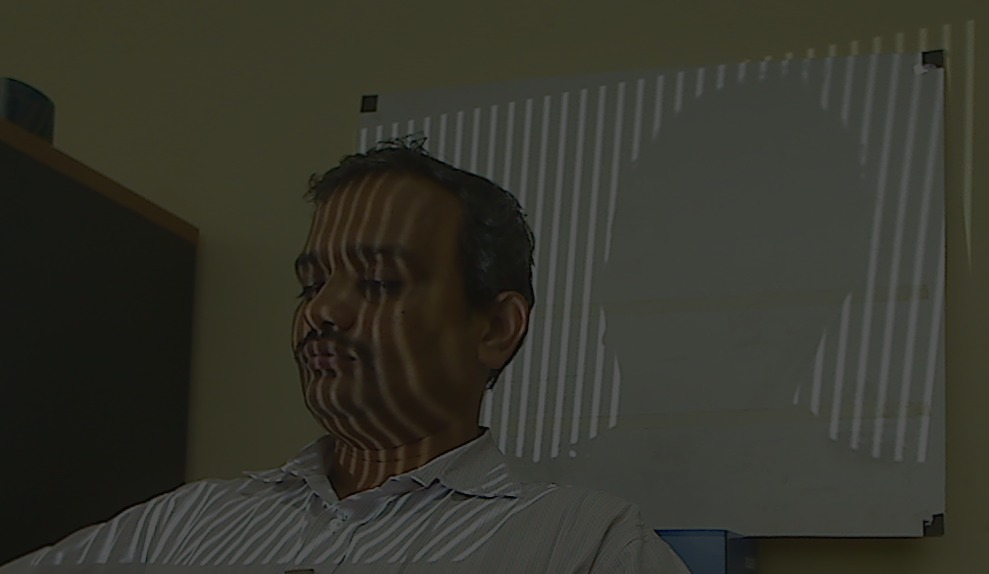}} & 
\subfloat[]{\includegraphics[width=4cm,height=4cm]{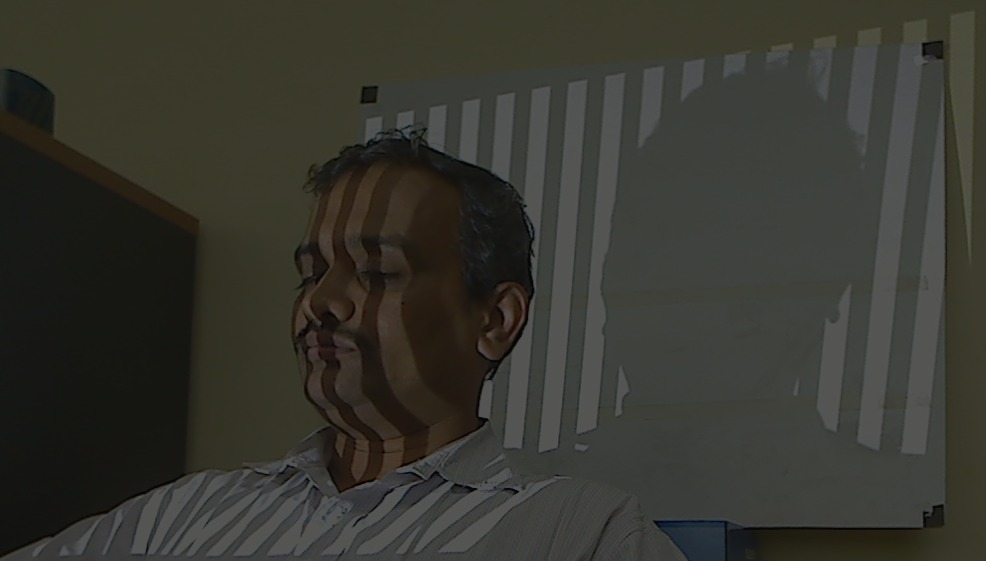}} \\
\end{tabular}
\end{tabularx}
\caption{Captured fringe and binary patterns}
\end{figure}

\subsubsection{Phase wrapping module}
As already mentioned, periodic nature of sinusoidal leads to
a recovered value of phase which repeats itself after period of
$2\pi$. This makes correspondence ambiguous since multiple
points with common phase value exists. This situation is
shown in (2) where computed phase wraps-up after every $2\pi$
interval. Hence, the computed phase is called \textit{wrapped
phase}. Figure 4 shows wrapped phase across the scene of
interest.
\begin{equation}
\phi=tan^{-1}\bigg[\frac{\sqrt[2]{3}(I_1-I_3)}{2I_2-I_1-I_3}\bigg]
\end{equation}
where, $-\pi\leq\phi\leq\pi$ 

\begin{figure}[ht]
\def\tabularxcolumn#1{m{#1}}
\begin{tabularx}{\linewidth}{@{}cXX@{}}
\begin{tabular}{c c}
\subfloat[Vertical wrapped phase]{\includegraphics[width=4cm,height=4cm]{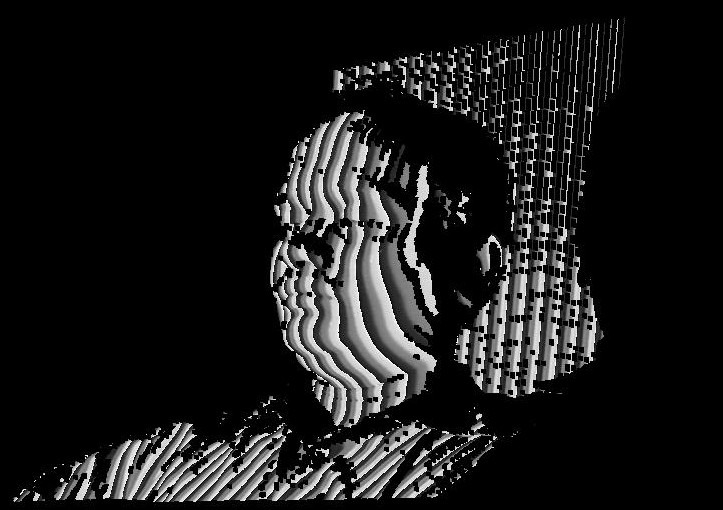}} &
\subfloat[Horizontal wrapped phase]{\includegraphics[width=4cm,height=4cm]{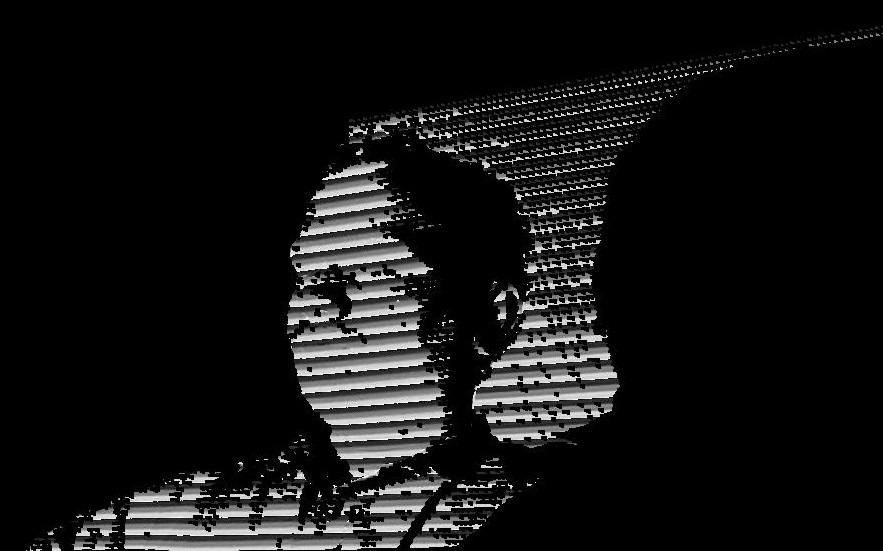}}\\
\end{tabular}
\end{tabularx}
\caption{Computed wrapped phase}
\label{fig:wrapped_phase}
\end{figure}

\subsubsection{Phase unwrapping module}
Wrapped phase computed in (2) repeats its value after every 2$\pi$ interval thereby making discrimination between pixels from different intervals non-trivial. This problem can be resolved by assigning
unique period number to each sinusoidal cycle(or interval). As described earlier, binary coded patterns with width of a bit-plane equal to the width of a fringe(or one sinusoidal cycle) are used for this purpose. Therefore the
true phase of incident signal at any camera pixel can be
represented as in equation 3. 
Since wrapped phase is unrolled/unwrapped by this process,
computed phase in this process is called \textit{unwrapped phase}.
Figure 5 shows the corresponding vertical and horizontal unwrapped phase.
\begin{equation}
\delta(x,y)=\phi(x,y)+2\pi*C(x,y)
\end{equation}
\noindent
where, $\delta(x,y)$ represents the unwrapped phase at pixel \textit{(x,y)} and \textit{C(x,y)} represents the decoded binary code at pixel \textit{(x,y)}
\begin{figure}[ht]
\def\tabularxcolumn#1{m{#1}}
\begin{tabularx}{\linewidth}{@{}cXX@{}}
\begin{tabular}{l r}
\subfloat[Vertical unwrapped phase]{\includegraphics[width=4cm,height=4cm]{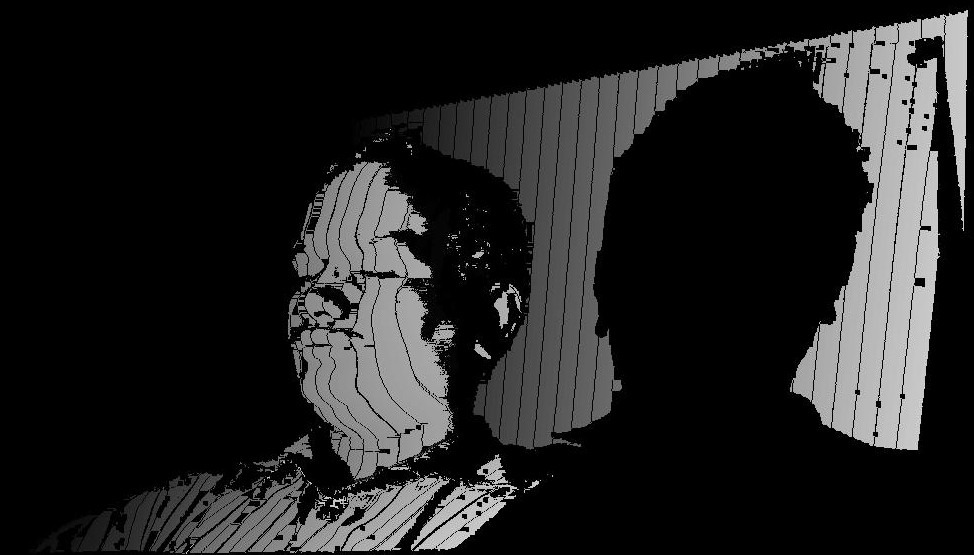}} &
\subfloat[Horizontal unwrapped phase]{\includegraphics[width=4cm,height=4cm]{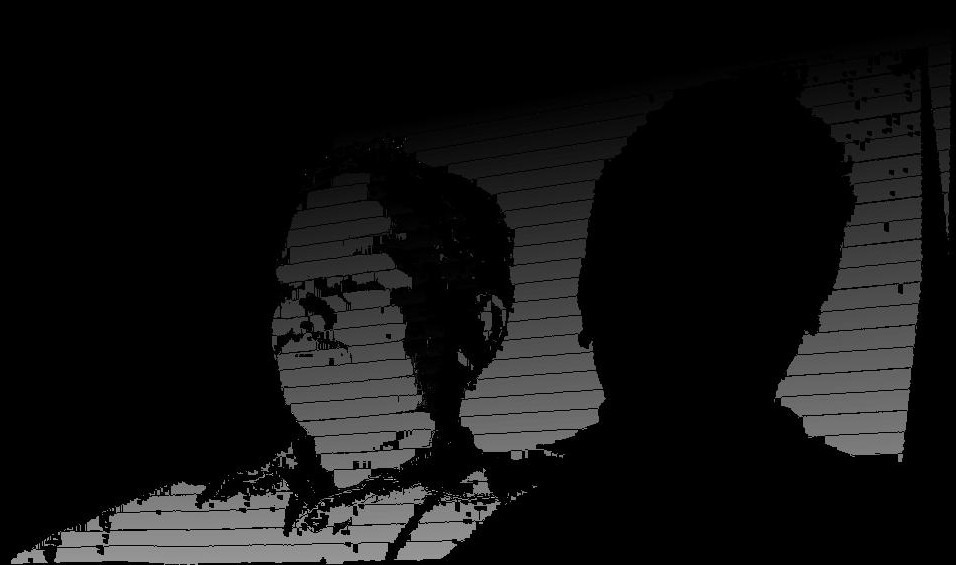}}\\
\end{tabular}
\end{tabularx}
\caption{Computed unwrapped phase}
\label{fig:unwrapped_phase}
\end{figure}

\indent Typically, thresholding techniques are used to recover
the codeword at each pixel from the captured images. Major
issue in accurate codeword extraction is faced at edges of the strips in these patterns. Practically, this region has a smooth
gradient instead of a hard edge, leading to ambiguity in codeword extraction in this region as mentioned in [16-17].

\subsubsection{Absolute phase computation}
Vertical unwrapped phase $\delta_v(x_c,y_c)$ gives projector X-coordinate($X_p$) whereas horizontal unwrapped phase $\delta_h(x_c,y_c)$ gives projector Y-coordinate($Y_p$) corresponding to camera-pixel $(x_c,y_c)$ for a fringe width $w_{fringe}$. Combining this information gives projector
pixel coordinates corresponding to camera coordinates.
Equation (4) explains this camera-to-projector coordinate
mapping. Figure 6 shows one example of estimated stereo-correspondence.
\begin{equation}
\begin{aligned}
& X_p=\lfloor w_{fringe}*\big(\frac{\delta_v(x_c,y_c)}{2\pi}\big) \rfloor \\ 
& Y_p=\lfloor w_{fringe}*\big(\frac{\delta_h(x_c,y_c)}{2\pi}\big) \rfloor
\end{aligned}
\end{equation}
\begin{figure}[htbp]
\def\tabularxcolumn#1{m{#1}}
\begin{tabularx}{\linewidth}{@{}cXX@{}}
\begin{tabular}{c c}
\subfloat[Camera image]{\includegraphics[width=4cm,height=4cm]{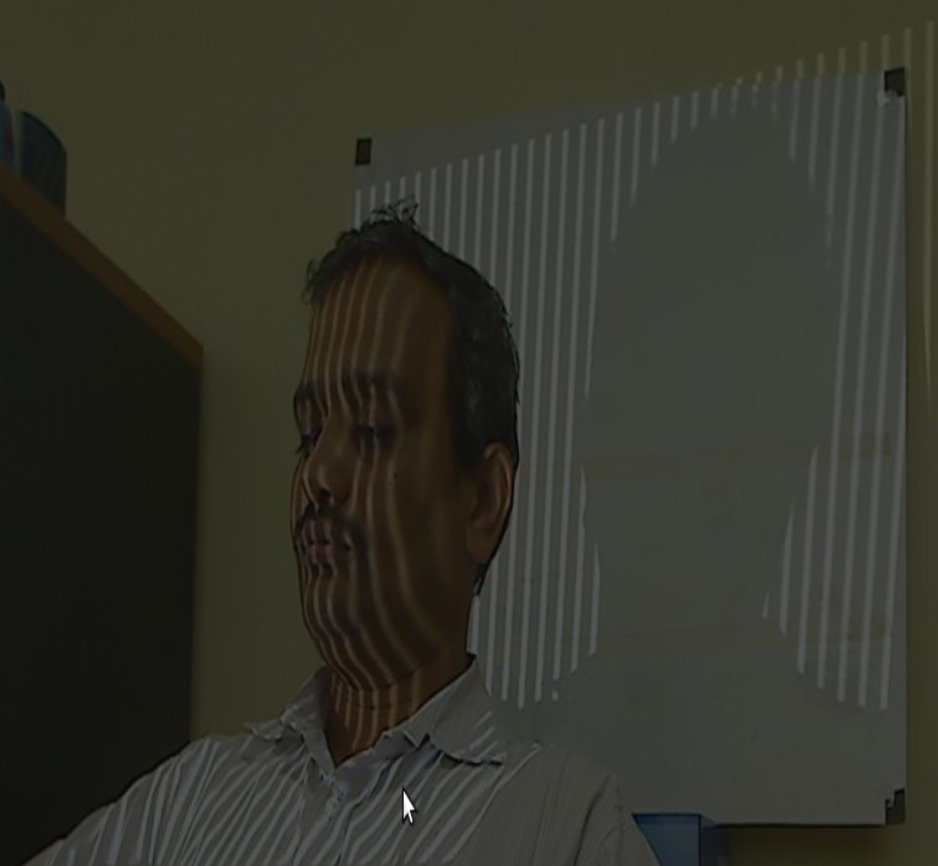}} &
\subfloat[Projector image]{\includegraphics[width=4cm,height=4cm]{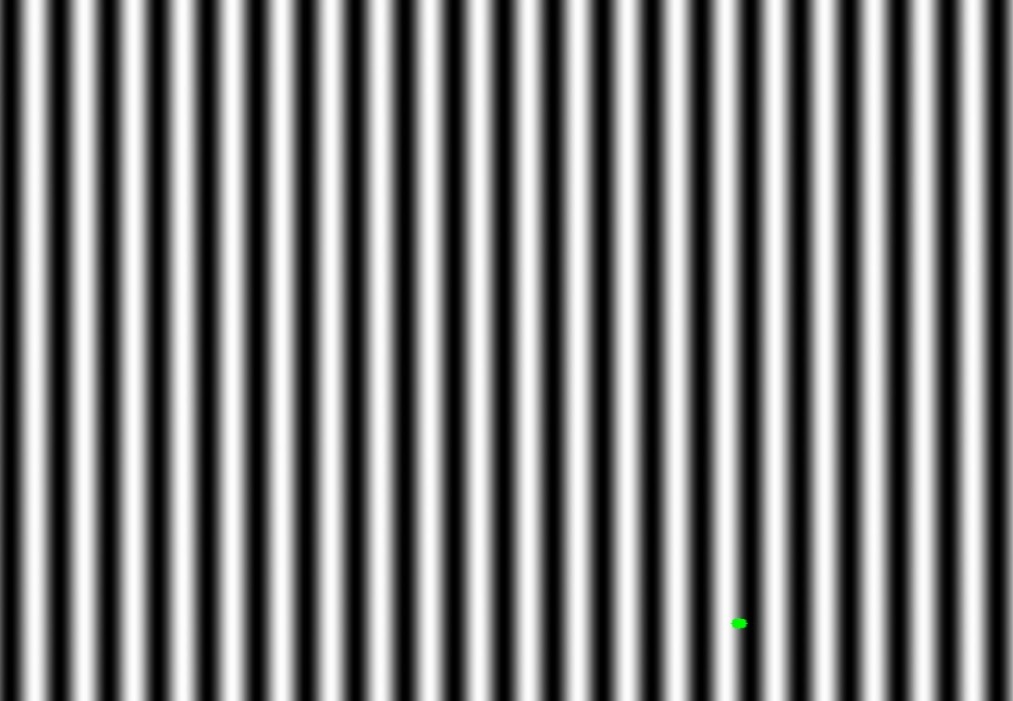}} 
\end{tabular}
\end{tabularx}
\caption{Stereo correspondence between camera and projector: \textit{green} spot in (b) corresponds to selected point(cursor) in (a)}
\label{fig:estimated_correspondence}
\end{figure}

\subsection{Triangulation}
System calibration parameters and camera-projector pixel-to-pixel correspondence information can be used to compute the
ray-ray intersections [18]. Solution to these equations will
give the 3D coordinates of real world point with respect to
world coordinate system. Figure 7 shows a example of 3D-
reconstructions obtained after solving these equation.
However, non-linear response of projector to input voltage
was found to be adding \textit{waviness} in the 3D reconstruction.
This effect was also observed in [19].
\begin{figure}
\def\tabularxcolumn#1{m{#1}}
\begin{tabularx}{\linewidth}{@{}cXX@{}}
\begin{tabular}{c c}
\subfloat[2D face image]{\includegraphics[width=4cm,height=4cm]{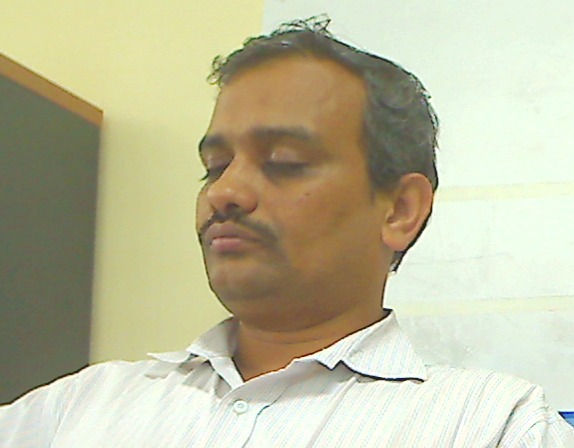}} &
\subfloat[3D view]{\includegraphics[width=4cm,height=4cm]{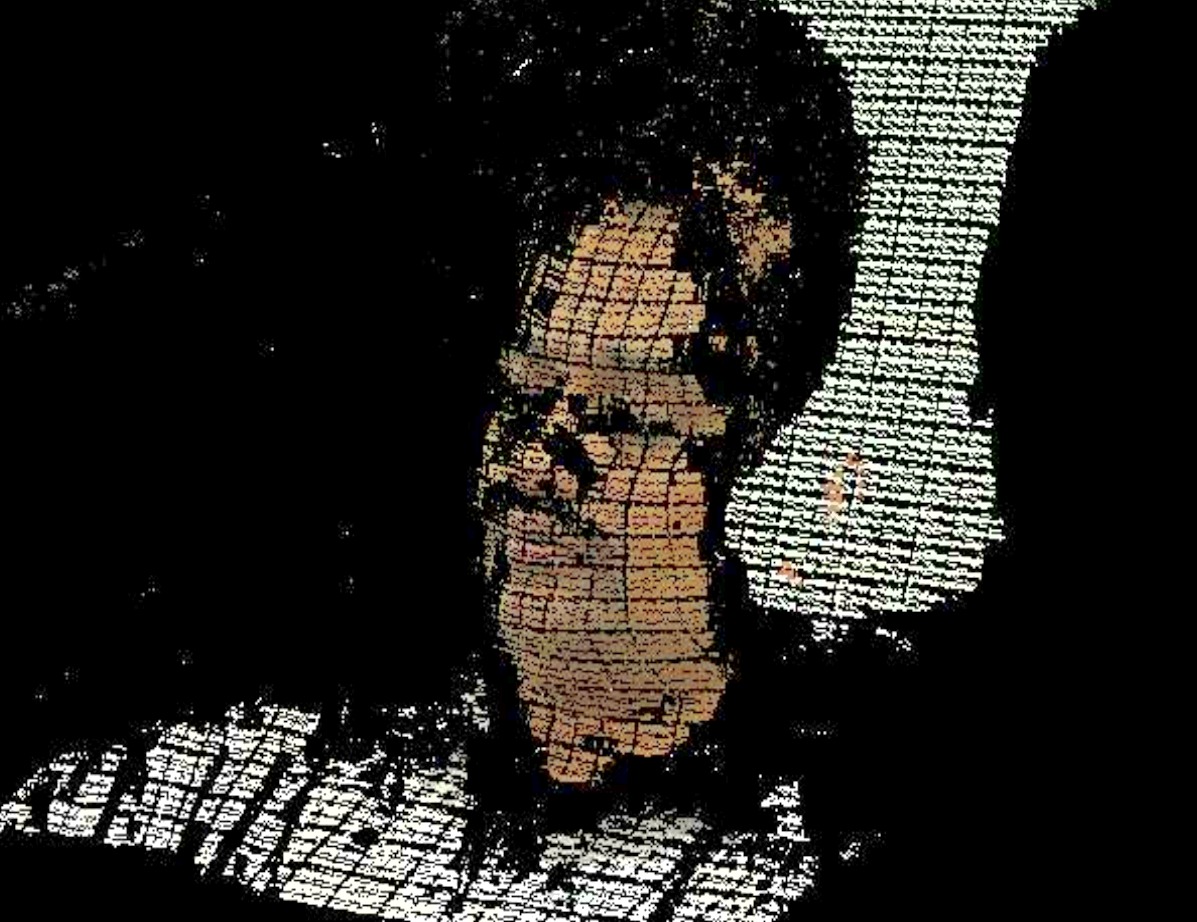}} 
\end{tabular}
\end{tabularx}
\caption{3D scan of a human face}
\end{figure}

\section{Accuracy Evaluation}
To evaluate measurement accuracy of system, 3D-distance
between selected feature-points in a planer checkerboard
were measured and compared against their true values. The
distance between camera-projector base-line and measurement object was decided based on common depth-of-field of camera-projector system such that acceptably sharper details of projected patterns can be acquired. In our
case it was $\sim2.2m$. Figure 8 shows the system pose used for measurement experiment. 
\begin{figure}
\centering
\includegraphics[width=10cm,height=10cm]{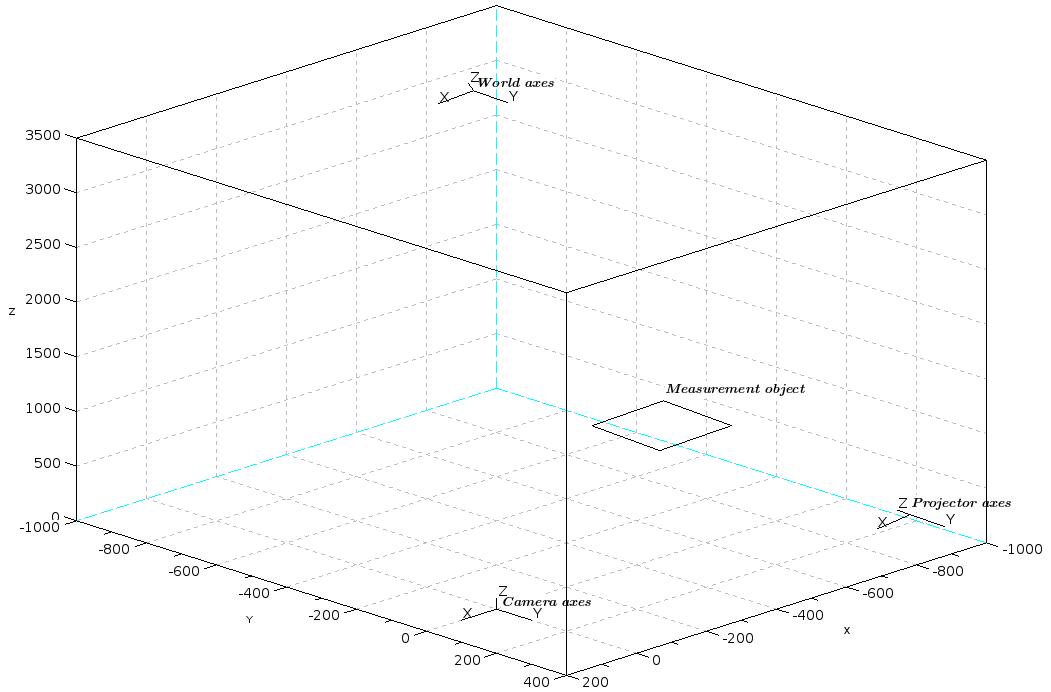}
\caption{System pose for 3D measurement experiment}
\end{figure}

Four inner-corners A,B,C,D (shown in Figure 9) were used to define lengths AB,BC,CD,DA,AC,BD. Measurement object was scanned 10 times to reduce the effect of non-systematic errors in the measurements. Average
percentage absolute relative error defined in equation (5) was
used as a measure of accuracy of the system.
\begin{figure}[hb]
\centering
\includegraphics[width=4.5cm,height=4cm]{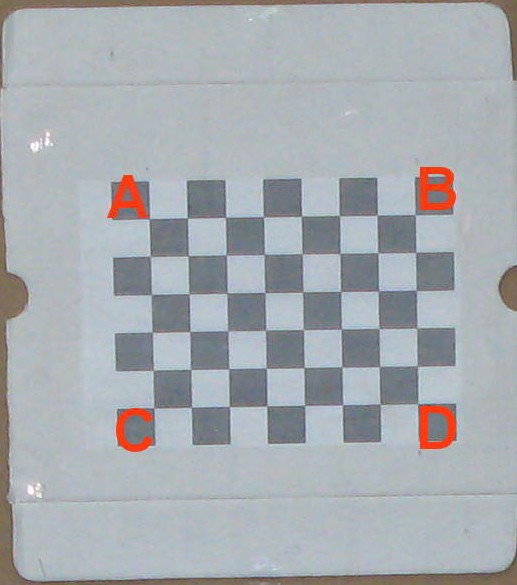}
\caption{Measurement object with corners A,B,C,D marked in \textit{red}}
\end{figure}
\begin{equation}
Accuracy=\frac{\sum_{i=1}^{N}\Bigg[\bigg(\frac{|actual_{i}-measured_{i}|}{actual_{i}}\bigg)*100\Bigg]}{N}
\end{equation}
where, $|x|$ denotes the absolute value of $x$ , $actual_i$ is the
true value for $i^{th}$ length measurement, $measured_i$ is the corresponding estimated value using 3D scanner data. $N$
denotes the total number of measurements.
To assess precision of the system, we determined average
of \% deviation of those 10 samples(6 measurements per
sample) with respect to their mean values as defined in equation (6). 
\begin{equation}
Precision=\frac{\sum_{p=1}^{vp}\Bigg[\frac{\sum_{i=1}^{vs_{p}}\Bigg[\bigg(\frac{|mean_{p}-sample_{i}|}{mean_{p}}\bigg)*100\Bigg]}{vs_{p}}\Bigg]}{vp}
\end{equation}
\noindent
where, $vp$ denotes total number of length measurements, $vs_{p}$ denotes total number of samples for $p^{th}$ length measurement, $mean_{p}$ and $sample_{i}$ denote the corresponding mean value and the $i^{th}$ sample respectively. 
Table II summarizes the observed results for measurement accuracy
and precision for our system with measurement object at a
distance of $\sim2.2m$ from camera-projector baseline.
\begin{table}[!t]
\caption{Measurement accuracy and precision of
developed coded phase-shift 3D-scanner}
\centering
\begin{tabular}{|c| |c|}
\hline
Metric & Value(in \%)\\
\hline
Measurement accuracy & 0.61\\
Precision & 0.29 \\ 
\hline
\end{tabular}
\end{table}

\section{Conclusion}
We have described a system for 3D scene reconstruction based on coded phase shift approach. Measurement accuracy and precision of system was evaluated and found to be within 1\% of true and mean measurements respectively. Developed system is designed to be experimental in nature allowing modification in various structured light and system calibration parameters. This will provide us a platform to investigate the problems related with system calibration
specifically the effect of relative pose of camera and projector used for calibration
on its accuracy(technically, \textit{sensor-planning}). In addition, it will allow us to study the
problem of projector-camera system non-linearity which is resulting in waviness in the 3D scan results. Recently, [20-21] have reported studies on effect of global illumination and projector defocus on accuracy of binary coded and phase shifting algorithms for stereo correspondence. These works have motivated our ongoing investigation of these issues in order to establish objective criteria for selecting a particular spatial frequency for binary coded and phase shifting patterns. This will reduce requirement of manual tweaking of system parameters to get optimal 3D measurement accuracy. Further, we have planned to extend the system to be able to do accurate 360 degree scans of the object.


\section*{Acknowledgment}
Authors would like to thank Computer Division,Bhabha
Atomic Research Centre technical staff and administration
for providing them support and facilities to pursue this work.



%

\begin{figure}[ht]
\includegraphics[width=8cm,height=11.5cm]{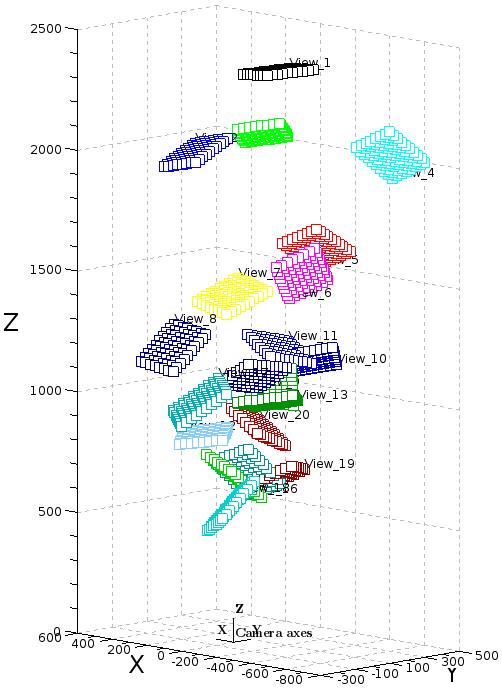}
\caption{Views used for camera calibration}
\end{figure}

\begin{figure}[ht]
\includegraphics[width=8cm,height=11.5cm]{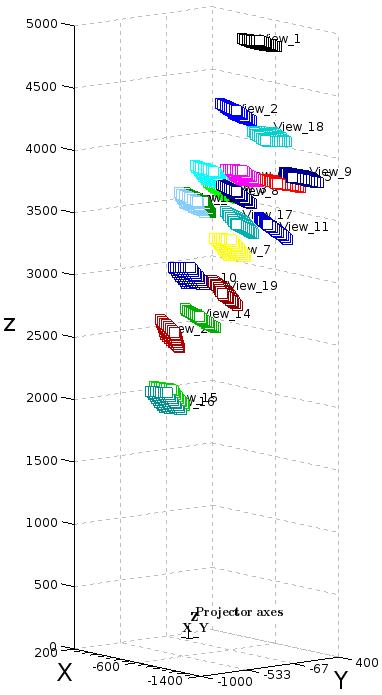}
\caption{Views used for projector calibration}
\end{figure}

\end{document}